\batchmode
\makeatletter
\def\input@path{{/home/timme/paper/tFROMd/submit180806//}}
\makeatother
\documentclass[letterpaper,american,prl, showpacs, floatfix, preprint]{revtex4}
\usepackage{times}
\usepackage[T1]{fontenc}
\usepackage[latin1]{inputenc}
\usepackage{graphicx}
\usepackage{amssymb}

\makeatletter

\providecommand{\boldsymbol}[1]{\mbox{\boldmath $#1$}}

\usepackage{float}
\usepackage{amsmath}
\usepackage{amssymb}

\usepackage{babel}

\usepackage{babel}
\makeatother
\begin{document}

\title{Revealing Network Connectivity From Dynamics}

\author{Marc Timme }

\affiliation{Center for Applied Mathematics, Theoretical \& Applied Mechanics,
Kimball Hall, Cornell University, Ithaca, NY 14853, USA}

\affiliation{Network Dynamics \emph{}Group\emph{,} Max Planck Institute for Dynamics
\& Self-Organization, }

\affiliation{Bernstein Center for Computational Neuroscience, Bunsenstr. 10, 37073
Göttingen, Germany}

\date{Fri Aug 18 11:57:46 EDT 2006
}

\begin{abstract}
We present a method to infer network connectivity from collective
dynamics in networks of synchronizing phase oscillators. We study
the long-term stationary response to temporally constant driving.
For a given driving condition, measuring the phase differences and
the collective frequency reveals information about how the oscillators
are interconnected. Sufficiently many repetitions for different driving
conditions yield the entire network connectivity from measuring the
dynamics only. For sparsely connected networks we obtain good predictions
of the actual connectivity even for formally under-determined problems.
\end{abstract}


\maketitle
Synchronization of networks of coupled units is an ubiquitous phenomenon
in nature. It occurs on very different temporal and spatial scales
in a variety of systems as different as Josephson junction arrays
and networks of neurons in the brain \cite{Strogatz,Pikovsky,Stewart}.
A central issue in current multi-disciplinary research is to understand
the relations between network structure and network dynamics. Given
an idealized model of the dynamics of the individual units and of
their interactions, what can we tell about features of the collective
dynamics depending on the network connectivity, say a regular lattice,
a random network or some more intricately connected network \cite{Bornholdt,Strogatz,Stewart}?
For many biological systems, such as networks of neurons, interacting
proteins or genes, and ecological foodwebs \cite{Chklovskii,Alon,Drossel,Berlow,Gardner,RMM},
however, important aspects of the network structure are largely unknown
such that inverse methods may prove useful. It would thus be desirable
to answer the reverse question: Can we infer information about the
connectivity of a network from controlled measurements of its dynamics?

Here we follow this novel perspective for synchronizing phase oscillators
that interact on networks of general connectivity. When driving one
or more oscillators, the measured phase dynamics reveals information
about the specific connectivity. We demonstrate that and how, given
a network of $N$ oscillators, each experiment (consisting of driving
and measuring) provides $N$ restrictions onto the network connectivity
that is defined by $N^{2}$ coupling strengths. Exploiting this, we
reveal the entire network \emph{connectivity} by repeatedly performing
measurements of the \emph{dynamics} only, under $N$ independent driving
conditions. Furthermore, assuming that real networks are substantially
more sparsely connected than all-to-all, we extend the method to reliably
predict the entire connectivity of the network even by a number of
experiments that is much smaller than the number of units in the network.

We consider a system of $N$ Kuramoto oscillators, a paradigmatic
model that has been successfully used to understand collective dynamical
phenomena in engineering, physics, chemistry, biology and medicine
\cite{Kuramoto,StrogatzKuramoto,Acebron,Maistrenko,Tass,Arenas}.
The oscillators are coupled on a directed network of unknown connectivity
with their dynamics satisfying \begin{equation}
\dot{\phi_{i}}=\omega_{i,0}+\sum_{j=1}^{N}J_{ij}\sin(\phi_{j}-\phi_{i})+I_{i,m}\label{eq:FullDynamics}\end{equation}
where $\phi_{i}(t)$ is the phase of oscillator $i$ at time $t$,
$\omega_{i,0}$ its natural frequency and $J_{ij}$ the coupling strength
from oscillator $j$ to $i$ ($J_{ij}=0$ if this connection is absent).
The quantity $I_{i,m}$ defines the strength of an external signal
to oscillator $i$ for driving condition $m$; it is identically zero,
$I_{i,0}\equiv0$, if the network is not driven. We define the in-degree
$k_{i}:=|\{ J_{ij}\neq0\,|\, j\in\{1,\ldots,N\}\}|$ as the number
of incoming links to oscillator $i.$

Consider the stationary dynamics on a phase locked attractor that
is close to in-phase synchrony and thus satisfies $\phi_{i}(t)-\phi_{j}(t)=d_{ij}$
where the $d_{ij}$ , $|d_{ij}|\ll1$, are constant in time. Networks
satisfying $J_{ij}\geq0$ and $|\omega_{i,0}-\omega_{j,0}|$ sufficiently
small for all $i,j$ exhibit such a stable phase-locked state close
to synchrony. The phase-locked condition for the free (undriven) dynamics
reads \begin{equation}
\Omega_{0}=\omega_{i,0}+\sum_{j=1}^{N}J_{ij}\sin(\phi_{j,0}-\phi_{i,0})\label{eq:PhaseLocked0}\end{equation}
where $\Omega_{0}$ is the collective frequency. %
\begin{figure}
\begin{center}\includegraphics[%
  width=80mm,
  keepaspectratio]{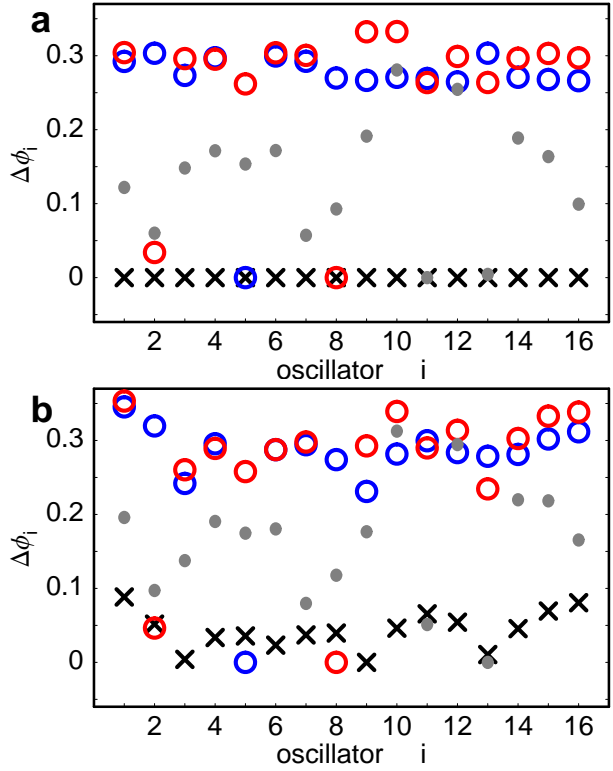}\end{center}

\caption{(color) Driving induces phase patterns, implicitly defined by (\ref{eq:PhaseLockedm}).
The network has $N=16$ oscillators, each connected with a constant
coupling strength $J_{ij}=1/k_{i}$ to $k_{i}\equiv8$ randomly selected
others ($J_{ij}=0$ otherwise). (a) Homogeneous frequencies, $\omega_{i}\equiv1$;
(b) inhomogeneous random frequencies $\omega_{i}\in[1,1+\Delta\omega]$,
$\Delta\omega=0.1$. Both panels display the phase differences $\Delta\phi_{i}:=\max_{j}\{\phi_{j}\}-\phi_{i}$
in the phase-locked states versus $i$. The responses to three different
driving conditions, (blue $\boldsymbol{{\bigcirc}}$) one oscillator
$i=5$ driven, $I_{5,1}=0.3$; (red $\boldsymbol{{\bigcirc}}$) two
oscillators $i\in\{2,8\}$ driven, $I_{2,2}=I_{8,2}=0.3$; (grey $\boldsymbol{{\bullet}}$)
all oscillators driven by a signal of random strength $I_{i,3}\in[0,0.3]$
are shown along with the undriven dynamics ($\boldsymbol{{\times}}$).
\label{cap:PhasePattern}}
\end{figure}

For synchronizing systems, commonly only one or a few scalar quantities
(such as one complex order parameter) are computed from measured dynamical
data (such as the oscillators' phases), often resulting in a statistically
accurate description of the overall network dynamics. Here we take
a complementary approach and seek a more detailed description of the
network dynamics in order to exploit this information for revealing
network connectivity.

We drive one or more oscillators in the network by temporally constant
input signals $I_{i,m}$, $i\in\{1,\ldots,N\}$ that can be positive,
negative or zero (meaning that oscillator $i$ is not driven). Such
inputs effectively change their natural frequencies. Keeping the signal
strengths sufficiently small, we structurally perturb the phase-locked
state such that it stays phase-locked and close to the original (cf.
Fig. \ref{cap:PhasePattern}). Such a driving signal results in a
phase pattern of the entire network that depends on the details of
the connectivity of that network as well as on the driving signal
itself \cite{Zanette,Kori,Denker,HMO,BlankBunimovich,EPL}. The perturbed
phase-locked state satisfies \begin{equation}
\Omega_{m}=\omega_{i,0}+\sum_{j=1}^{N}J_{ij}\sin(\phi_{j,m}-\phi_{i,m})+I_{i,m}\label{eq:PhaseLockedm}\end{equation}
for all $i\in\{1,\ldots,N\}$ where now $\Omega_{m}$ is the new collective
frequency that has changed due to the driving and $\phi_{i,m}$ are
the stationary phases in response to the driving. 

Now take the differences between the phase-locked conditions for the
driven and the undriven system, \begin{equation}
D_{i,m}=\sum_{j=1}^{N}J_{ij}\left[\sin(\phi_{j,m}-\phi_{i,m})-\sin(\phi_{j,0}-\phi_{i,0})\right]\label{eq:PhaseLockedDifference}\end{equation}
where $D_{i,m}:=\Omega_{m}-\Omega_{0}-I_{i,m}$. For sufficiently
small structural perturbations we approximate $\sin(x)=x+\mathcal{{O}}(x^{2})$
and abbreviate the phase shifts $\theta_{j,m}:=\phi_{j,m}-\phi_{j,0}$
yielding\begin{equation}
D_{i,m}=\sum_{j=1}^{N}\hat{{J}}_{ij}\theta_{j,m}\,\label{eq:DiffMatrix}\end{equation}
where $\hat{J}$ is the $N\times N$ Laplacian matrix is given by\begin{equation}
\hat{{J}}_{ij}=\left\{ \begin{array}{lll}
J_{ij} & \textrm{for} & i\neq j\\
-\sum_{k,k\neq j}J_{ik} & \textrm{for} & i=j\end{array}\right..\label{eq:Jhat}\end{equation}

Given one driving condition $m$, we measure $N-1$ independent phase
shifts $\theta_{i,m}$ and one collective frequency $\Omega_{m}$
to obtain $N$ linearized equations (\ref{eq:DiffMatrix}) that restrict
the $N^{2}$ dimensional space of all possible network connectivities
$(J_{ij})_{i,j\in\{1,\ldots,N\}}$. This is the maximum number of
restrictions one can deduce from one experiment. As a consequence,
from measurements under linearly independent driving conditions, we
obtain more and more information about the connectivity: After performing
$M$ experiments \cite{outliers} the space of networks is restricted
by $MN$ equations \begin{equation}
D=\hat{J}\theta\label{eq:MatrixEquation}\end{equation}
where $\theta=(\theta_{i,m})_{i\in\{1,\ldots,N\},m\in\{1,\ldots,M\}}$
is the $N\times M$ matrix of column vectors of phase differences
for each experiment $m$ and, analogously, $D=(D_{i,m})_{i,m}$ is
the $N\times M$ matrix of the effective frequency offsets. Thus we
are left with an $(N-M)N$-dimensional family of possible networks
that are consistent with the $M$ measured data sets. In particular,
this implies that after $M=N$ experiments the network connectivity
is specified completely as given by $\hat{J}=D\theta^{-1}$. We thus
find the network \emph{connectivity} from measuring the response \emph{dynamics}
only, see, e.g. Fig. 2. 

\begin{figure}
\begin{center}\includegraphics[%
  width=72mm,
  keepaspectratio]{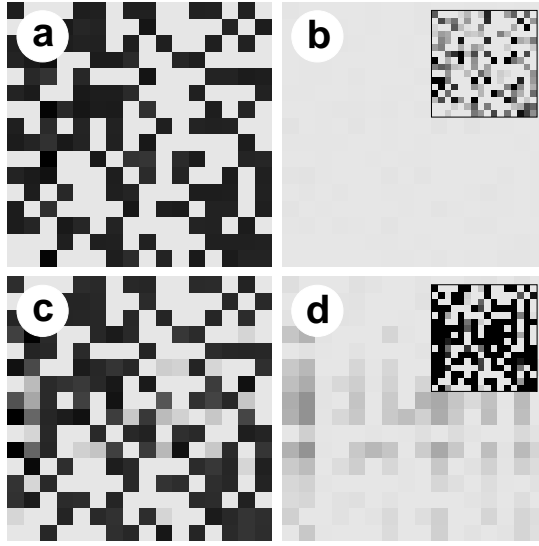}\end{center}

\caption{Inferring connectivity from measuring response dynamics. $M=N=16$
experiments \cite{DrivingSignals}. (a) Connectivity of the network
of Fig 1a as obtained using Eqs. (\ref{eq:DiffMatrix}--\ref{eq:MatrixEquation}).
The matrix of connection strengths $J_{ij}$ is gray-coded from light
gray ($J_{ij}=0$) to black ($J_{ij}=\max_{i',j'}\{ J_{i'j'}\}$).
(b) Element-wise absolute difference $|J_{ij}^{\textrm{original}}-J_{ij}^{\textrm{derived }}|$,
plotted on the same scale as (a). Inset shows magnified difference
$100\times|J_{ij}^{\textrm{original}}-J_{ij}^{\textrm{derived }}|$
with a cutoff at unity (black). Panels (c) and (d) are analogous to
(a) and (b) for the network with inhomogeneous frequencies of Fig.
1b. \label{cap:Nmeasurements}}
\end{figure}

This direct method in principle works for networks with homogeneous
as well as with inhomogeneous frequencies \cite{NonsystematicQuality}.
The method is capable of revealing not only which links are present
and which are absent but also gives a good quantitative estimate of
the actual link strengths $J_{ij}$. It has, however, also some drawbacks.
The problem of solving (\ref{eq:MatrixEquation}) can be ill-conditioned
in the sense that the ratio of the largest and smallest singular value
of $\theta^{\textrm{T}}$ is large, leading to low-quality reconstruction.
Moreover, the direct method might become impractical when studying
real-world networks which often consist of a large number $N$ of
units and thus would require a large number $M=N$ of (possibly costly)
experiments.

Can we obtain the connectivity more efficiently, even with $M<N$
experiments? In many networks, such as networks of neurons in the
brain, a substantial number of potential links are \emph{not} present:
each node $i$ typically has a number $k_{i}\ll N$ of links. Here
we exploit this fact and look for that connectivity matrix $J$ that
has the least number of links (maximum number of $J_{ij}=0$) but
is still consistent with all $M$ measured data sets.

\begin{figure}
\begin{center}\includegraphics[%
  width=72mm,
  keepaspectratio]{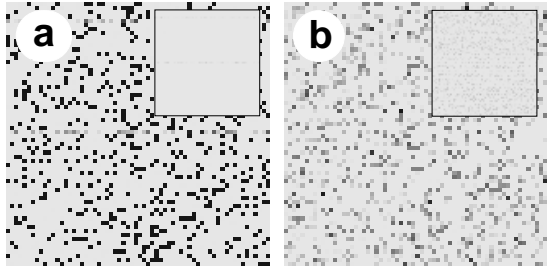}\end{center}

\caption{Revealing connectivity with $M<N$ measurements. Network ($N=64$,
$k=10$, $\Delta\omega=0$) reconstructed by minimizing the 1-norm,
(a) $M=38$, (b) $M=24$. The insets show the element-wise absolute
differences to the original network.\label{cap:MlessthanNmeasurements}}
\end{figure}

To achieve this goal we use the constraints (\ref{eq:MatrixEquation})
to parameterize the family of admissible matrices by $(N-M)N$ real
parameters $P_{ij}$, $i\in\{1,\ldots,N\}$, $j\in\{ M+1,\ldots,N\}$
in a standard way using a singular value decomposition of $\theta^{\textrm{T}}=USV^{\textrm{T}}$
where the $M\times N$ matrix $S$ contains the singular values on
the diagonal, $S_{ij}=\delta_{ij}\sigma_{i}\geq0$. We rewrite the
set of all coupling matrices $\hat{J}=DU\tilde{S}V^{\textrm{T}}+PV$
, setting $P_{ij}=0$ for all $j\leq M$ and $\tilde{S}_{ij}=\delta_{ij}/\sigma_{i}$
if $\sigma_{i}>10^{-4}$ and $\tilde{S}_{ii}=0$ if $\sigma_{i}\leq10^{-4}$
. Finally, we minimize the 1-norms of the row vectors of $\hat{J}$
(input coupling strengths)\begin{equation}
\left\Vert \hat{J}_{i}\right\Vert _{1}:=\sum_{j=1;j\neq i}^{N}\left|J_{ij}\right|\label{eq:oneNorm}\end{equation}
 with respect to the parameters $P$, separately for all oscillators
$i$. By this method we find the network $J$ that is closest to the
origin $J=0$, which in 1-norm is one with a minimal number of incoming
links (maximal number of zero entries) \cite{ConvexOptimization};
thus we find the sparsest of all networks satisfying the measurement
data. Reasonably good reconstructions can already be obtained with
the number of experiments $M$ being substantially smaller than $N$,
as illustrated in Fig. \ref{cap:MlessthanNmeasurements}. 

\begin{figure}[th]
\begin{center}\includegraphics[%
  width=120mm,
  keepaspectratio]{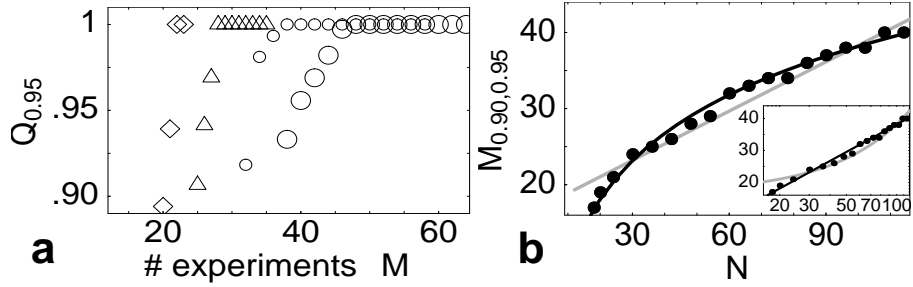}\end{center}

\caption{Quality of reconstruction and required number of experiments. (a)
Quality of reconstruction ($\alpha=0.95$) displayed for $k=10$ and
$N=24$ ({\LARGE $\diamond$}), $N=36$ ($\triangle$), $N=66$ ($\circ$),
$N=96$ ($\bigcirc$) . (b) Minimum number of experiments required
($q=0.90$, $\alpha=0.95$) versus network size $N$ with best linear
and logarithmic fits (gray and black solid lines). Inset shows same
data with $N$ on logarithmic scale. }

\label{cap:QualityM}
\end{figure}

How reliable is such a reconstruction? This depends on the details
of the network connectivity and the realization of driving. We did
a case study for random networks of different sizes $N$, where each
oscillator receives input connections from $k_{i}\equiv k<N$ randomly
chosen others. Using $J_{\textrm{max }}=\max_{i',j'}\left\{ \left|J_{i'j'}^{\textrm{derived}}\right|,\left|J_{i'j'}^{\textrm{original}}\right|\right\} $,
define the element-wise relative difference as \begin{equation}
\Delta J_{ij}:=\left.J_{\textrm{max}}^{-1}\left|J_{ij}^{\textrm{derived}}-J_{ij}^{\textrm{original}}\right|\right/2\label{eq:deltaJ}\end{equation}
 such that $\Delta J_{ij}\in[0,1]$. After $M$ experiments, the quality
of reconstruction is defined as the fraction\begin{equation}
Q_{\alpha}(M):=\frac{1}{N^{2}}\sum_{i,j}H\left((1-\alpha)-\Delta J_{ij}\right)\,\in[0,1]\label{eq:Qualilty}\end{equation}
 of coupling strengths which are considered correct. Here $\alpha\leq1$
is the required accuracy of the coupling strengths and $H$ the Heaviside
step function, $H(x)=1$ for $x\geq0$, $H(x)=0$ for $x<0$. Typically,
the quality of reconstruction increases with $M$ (but depends also
on the realizations of the experiments), becoming close to one already
for $M$ substantially smaller than $N$, see Fig. \ref{cap:QualityM}a.
We furthermore evaluated the minimum number of experiments \begin{equation}
M_{q,\alpha}:=\min\{ M\,|\, Q_{\alpha}(M)\geq q\}\label{eq:Mrequired}\end{equation}
required for accurate reconstruction on quality level $q$. Figure
\ref{cap:QualityM}b shows $M_{0.90,0.95}$ , the minimum number of
experiments required for having at least $q=90\%$ of the links accurate
in strength on a level of at least $\alpha=95\%$, as a function of
$N$. The numerics suggests that $M_{q,\alpha}$ generally scales
sub-linearly (presumably logarithmically) with network size $N$ for
reasonable $0<1-\alpha\ll1$ and $0<1-q\ll1$. In particular, it implies
that the connectivity of a network can be revealed reliably even if
$M$ is much smaller than the network size $N$.

Recently, the response of synchronizing phase oscillators to different
kinds of driving signals has been studied for random networks and
lattices \cite{Denker,HMO,Kori,Zanette}. In the present study we
took advantage of the fact that in response to controlled driving
(cf. also \cite{BlankBunimovich,EPL}) the dynamics induced may critically
depend on the network connectivity (cf. also Fig. \ref{cap:PhasePattern}).
This is generally the case if the networks are strongly connected
\cite{stronglyconnected} but have otherwise arbitrary connectivities
(cf. Eq.~(\ref{eq:PhaseLockedm}) and \cite{EPL}). Thus information
about the \emph{connectivity} can be revealed from measuring the response
\emph{dynamics}. To achieve this, we exploited all available information
of the network dynamics (the $N-1$ independent phase differences
and the collective frequency) rather than only statistical information,
such as one complex order parameter. Interestingly, in a recent study,
Arenas et al. \cite{Arenas} also used more detailed information of
the dynamics and successfully inferred the hierarchical structure
of a network.

The method presented here not only identifies where links are present
and where they are absent but also gives a good estimate for the strength
of each connection. For networks with a substantial number of potential
links absent, we furthermore showed how to predict the connectivity
in a reliable way even by a number of experiments that is much smaller
than the network size. In fact, the numerical evaluation suggests
that the number of experiments required for faithful reconstruction
only scales sub-linearly with the network size. The relatively simple
yet efficient method presented here thus qualifies as potentially
practically useful also for real systems of moderate or larger size
where the number of measurement might be desired as small as possible.
An important question for future research is thus how to extend the
method presented here to networks of dynamical elements that are described
by more than one variable and that possibly do not synchronize but
organize into some other, more complicated, collective state.

The multidisciplinary research community studying networks has recently
seen significant progress towards understanding the implications of
structural features for network dynamics and function, in particular
in biological networks. Interesting examples \cite{Gardner,Alon,Chklovskii}
include (i) network motifs, small subnetworks that occur significantly
more often than in randomized networks, have been identified in a
variety of complex systems and might be designed for functionality;
(ii) a small part of a genetic pathway was successfully identified
based on expression profiling; (iii) neural wiring in the brain appears
to follow optimization rules. Together with such findings, our results
on synchronizing oscillator networks suggest a very promising future
direction of research: Methods similar to the one presented here should
on the one hand help to better understand structure-dynamics relations
from measuring perturbed, possibly complicated, stable dynamics; on
the other hand they could also help clarifying structural questions
in the first place, e.g., by identifying functionally meaningful parts
of a network.

I thank C. Kirst and S. Strogatz for helpful discussions. I acknowledge
financial support by the BMBF Germany via the BCCN Göttingen under
grant 01GQ0430 and by the Max Planck Society.


\begin{thebibliography}{14}
\bibitem{stronglyconnected}A directed network is strongly connected if there is a directed path
between any ordered pair of nodes.
\bibitem{DrivingSignals}Here and in the following, the driving signals are independent random
$I_{i,m}\in[0,\Delta I]$, where $\Delta I=0.2$.
\bibitem{NonsystematicQuality}The reconstruction quality depends not only on the degree of inhomogeneity,
but also, among other parameters, on the realization of it and of
the driving.
\bibitem{outliers}We discard outliers: they may originate from perturbations that create
(i) no structurally stable response or (ii) responses that are close
to linearly dependent on others.
\end{thebibliography}
\end{document}